# Assessing Trust in Construction AI-Powered Collaborative Robots using Structural Equation Modeling

**Newsha Emaminejad[1], Lisa Kath[2], and Reza Akhavian[3*]**

[1] Graduate Student, Department of Civil, Construction, and Environmental Engineering, San Diego State University, San Diego, CA 92182, USA, Email: nemaminejad8591@sdsu.edu
[2] Associate Professor, Department of Psychology, San Diego State University, San Diego, CA 92182, USA, Email: lisa.kath@sdsu.edu
[3] Associate Professor, Department of Civil, Construction, and Environmental Engineering, San Diego State University, San Diego, CA 92182, USA, Email: rakhavian@sdsu.edu
*Corresponding Author

**Abstract:**
This study aimed to investigate the key technical and psychological factors that impact the architecture, engineering, and construction (AEC) professionals' trust in collaborative robots (cobots) powered by artificial intelligence (AI). The study employed a nationwide survey of 600 AEC industry practitioners to gather in-depth responses and valuable insights into the future opportunities for promoting the adoption, cultivation, and training of a skilled workforce to leverage this technology effectively. A Structural Equation Modeling (SEM) analysis revealed that safety and reliability are significant factors for the adoption of AI-powered cobots in construction. Fear of being replaced resulting from the use of cobots can have a substantial effect on the mental health of the affected workers. A lower error rate in jobs involving cobots, safety measurements, and security of data collected by cobots from jobsites significantly impact reliability, while the transparency of cobots' inner workings can benefit accuracy, robustness, security, privacy, and communication, and results in higher levels of automation, all of which demonstrated as contributors to trust. The study's findings provide critical insights into the perceptions and experiences of AEC professionals towards adoption of cobots in construction and help project teams determine the adoption approach that aligns with the company's goals workers' welfare.

## 1. Introduction

Traditional construction production methods rely heavily on human workers, require substantial, arduous effort, and can pose safety and health risks to workers due to exposure to demanding physical work and dangerous substances. Instances of physical hazards that can lead to injuries or long-term damage to the body include being subjected to vibrations or loud noises, as well as exposure to chemical hazards such as vapors, dust, or fumes [1]. Although construction industry has attempted to improve productivity by using powered hand tools and more recently prefabrication, it is still lagging behind other industries in their embrace of automation. [2]. Additionally, the industry has been facing a labor shortage, which has further slowed progress toward increased productivity [3]. According to the Associated Builders and Contractors, the construction industry will have to attract around 546,000 extra workers in 2023, in addition to the regular hiring pace, to meet the labor demand. This shortfall in workers is due to multiple factors, including an aging workforce, a lack of interest among younger generations to enter the industry, and immigration policies that limit the availability of foreign workers [4].

Therefore, the industry needs to improve productivity and construction time while addressing workers' safety concerns and cost overruns. This can be achieved by developing new solutions to carry out labor-intensive tasks using artificial intelligence (AI) and robotics, thus freeing up workers to focus on more technical jobs, reducing the risk of injuries caused by physical hazards and exposure to harmful substances, improving productivity and efficiency, and reducing labor costs [5-7]. To automate repetitive and linear tasks such as bricklaying, demolishing, and welding, AI and robotics have shown great promise in the execution phase of a project [8, 9].

Studying collaborative robots, or "cobots" in the industrial applications is receiving increasing attention from both engineering and social science research communities [10]. These are robots designed to work alongside human workers, enhancing their capabilities and assisting them in performing their tasks [11, 12]. Construction tasks are often too complex for full automation and robots require collaboration with humans. Thus, cobots are gaining popularity in the AEC industry and are expected to play a significant role in the future of construction [13].

However, the widespread adoption of cobot applications on construction sites has been limited due to various factors, such as lack of research, high initial costs, technical complexity, health and safety concerns for human-robot collaboration, potential job displacement, and non-compliance with building regulations [14-16]. Furthermore, existing commercially available solutions for physically demanding tasks such as automated masonry [17-19], multi-functional robots such as Baubot, which can perform tasks such as milling, drilling, sanding, and laser marking [20] or rebar laying and tying [21, 22] may raise the concern of limiting the involvement of the human worker in the loop, which exacerbates the concerns associated with robots taking over workers jobs. To address these unique aspects of the construction industry, establishing trust between human workers and cobots is imperative. The level of trust that humans have in cobots is a key consideration that impacts the success and effectiveness of human-robot teams.

Insufficient trust can result in disuse, where individuals are unwilling to use the robots and do not recognize their abilities [23]. Conversely, excessive trust can lead to overreliance on the robots, potentially causing failures in critical situations. Therefore, it is essential to have a calibrated level of trust in collaborative robots for successful human-robot interaction at construction sites. This study aims to investigate the technical and psychological factors that may influence the establishment and reinforcement of trust among AEC professionals when working with AI-enabled cobots to enable major future work on trust calibrations.

## 1.1 Research Background

In previous steps of this research program, a comprehensive literature review as well as interviews with AEC practitioners were conducted [24-26], and several factors about gaining users' trust in robots were identified. As shown in Figure 1, a set of 13 factors influencing trust were established.

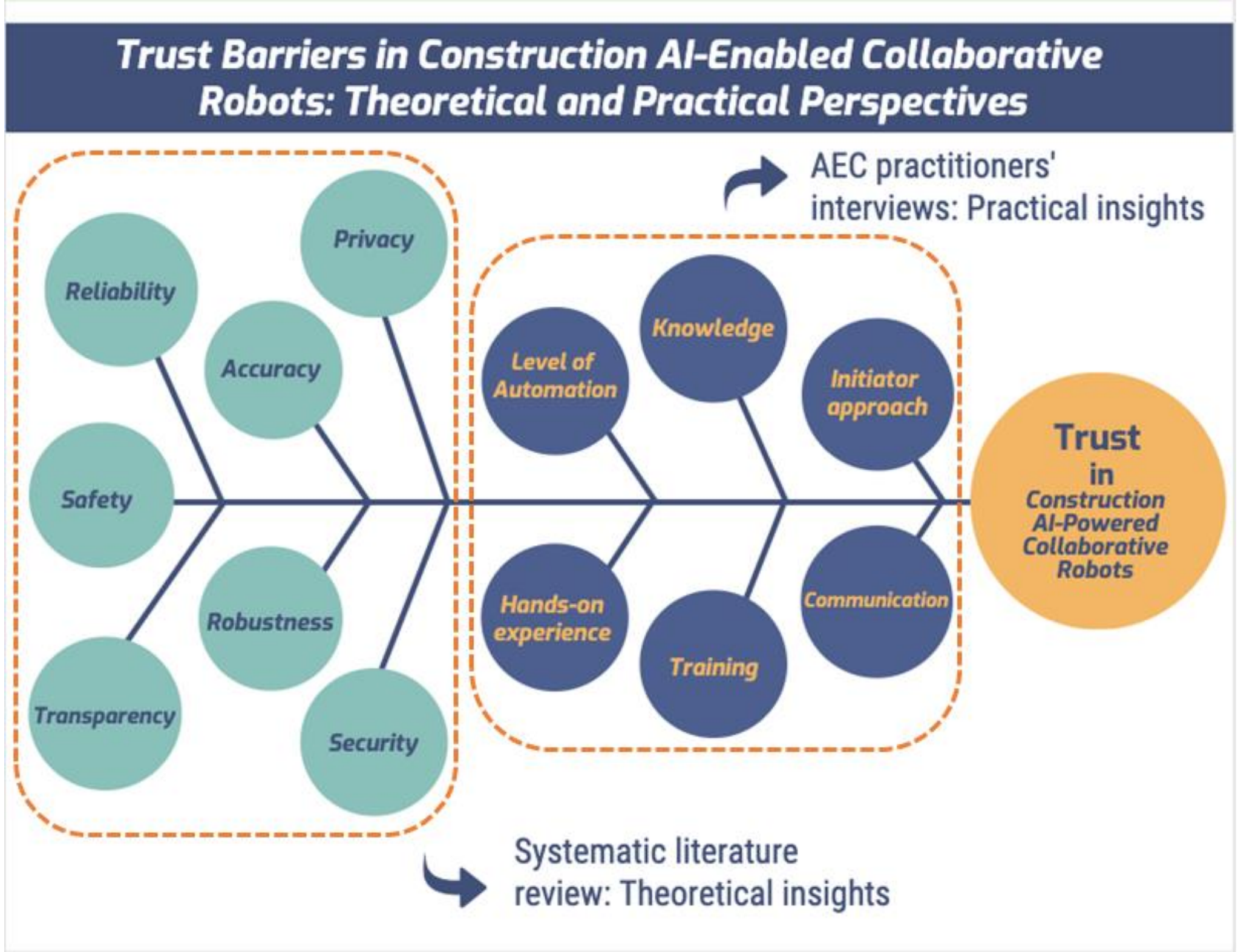

Figure 1. Trust barriers in construction AI-enabled collaborative robots (Cobots): Theoretical and practical perspectives

Among these, seven were directly derived from the initial stage of the current research, which involved a systematic literature review [24-26]. Subsequently, an additional six factors were defined based on insights gathered from interviews conducted with construction practitioners, constituting the second stage of the research [24-26].

1. **Reliability** in the context of cobots refers to their capacity to carry out their designated tasks consistently and predictably without experiencing any malfunctions, failures, or unexpected interruptions.

2. **Safety** in cobots encompasses the process of designing and deploying cobots to ensure the well-being and protection of human workers who collaborate with them. Since cobots are specifically designed to operate alongside humans, they must incorporate essential safety measures that minimize the risk of accidents, injuries, and damage to property.
3. **Transparency** in cobots pertains to their capacity to offer feedback regarding their operations and performance, enabling human workers to observe and comprehend the actions of the robot. This transparency ensures that external observers can easily understand how the system generated its outcomes.
4. **Robustness** in cobots refers to their capacity to function efficiently and consistently across diverse conditions, without being negatively impacted by variations in the environment or task demands.
5. **Accuracy** in cobots pertains to the capacity of these robots to execute their designated tasks with exceptional precision and minimal margin of error.
6. **Privacy** in cobots refers to the protection of personal information and data that may be collected or processed by collaborative robots.
7. **Security** of cobots entails safeguarding the cobot and its related systems and construction workflows against unauthorized access, malicious attacks, and other potential security risks.
8. **Level of automation** (LOA) in cobots pertains to the extent to which cobots can independently carry out their assigned tasks, without requiring human intervention or control. This level of automation is influenced by the complexity of the tasks cobots can handle and their ability to adjust to dynamic conditions or inputs.
9. **Hands-on experience** involves actively engaging workers with cobots in real-world scenarios. This encompasses tasks such as programming, operation, and maintenance across diverse construction environments. Hands-on experience fosters a profound comprehension of cobots' capabilities, constraints, and the wide range of potential applications they offer.

10. **Having general knowledge** about cobots involves possessing a basic understanding of cobots and their applications, even without hands-on experience. This includes being familiar with cobot features, components, and capabilities, as well as their potential benefits and limitations in various construction contexts. Several factors can influence the acquisition of this knowledge, such as:
    - Level of education: The educational background can shape individuals' understanding of cobots and provide a basis for comprehending their principles and applications.
    - Years of experience: The amount of time individuals' have spent working in the construction industry can contribute to their exposure to cobots and related technologies.
    - Familiarity with relevant technologies: Being acquainted with the technologies that enable cobots functionalities, can influence workers basic understanding of cobots.
    - Awareness about construction robotics: Being informed about the current use of robots in construction contributes to individuals' knowledge of cobots potential in projects.
11. **Training** involves participating in a structured program aimed at acquiring the skills to operate, program, and maintain cobots in a safe and efficient manner. The training curriculum generally encompasses key areas such as safety protocols, cobot programming, operation and maintenance procedures, troubleshooting techniques, and fault diagnosis. It often incorporates a blend of theoretical and practical instruction in simulated or real-world settings.
12. **Initiator approach** of trust in cobots is the individual or group that takes the initial action to establish trust in cobots within a specific context or environment. This initiator can be a union, a company's CEO, a manager, an engineer, or any other person accountable for the implementation and operation of these robots in a particular workplace or industry.

13. **Communication** with cobots entails the capability for humans to engage with cobots through diverse approaches, including voice commands, having two-way conversation, gestures, cobot size or shape, touchscreens, or alternative interfaces.

Factors 1-8 pertain to the cobot's system and configurations, 9-12 refer to user's perception, and 13 is related to the environmental conditions. In addition to defining these trust factors, the literature review and interviews gave birth to in several hypotheses about them. This paper describes the results of a nationwide survey to determine their relevance and applicability in practical perspectives (summarized in Figure 2), through testing the following hypotheses:

*H1. The perceived **reliability** of cobots is significantly affected by their **safety** value, including the reduction of potential harm and the overall safety improvement compared to non-cobot operations.*

*H2. The **safety** of cobots is significantly influenced by their **robustness** and ability to perceive and understand the environment.*

*H3. **Training** directly influences the perception of **safety** regarding cobots. In other words, well-trained individuals are expected to have a higher perception of safety regarding cobots compared to those who have received less or no training.*

*H4. The **level of automation** in cobots has a direct impact on how individuals perceive their **safety**. It suggests that as the level of automation increases, individuals' perception of safety regarding cobots will also increase.*

*H5. Sufficient **knowledge** about cobots, their capabilities, and limitations positively influences how they are embraced by workers and organizations.*

*H6. Individuals who have **firsthand experience** of working directly with cobots are more likely to embrace these robots compared to those who have not interacted with them in this manner.*

*H7.The **transparency** of cobots' inner workings and decision-making processes plays a significant role in determining the acceptable **level of automation**.*

*H7a. The level of **transparency** in the decision-making process of cobots is significantly correlated to their **communication** capabilities. When cobots have a transparent decision-making process, they can effectively convey their actions and intentions to users, resulting in clearer and more efficient communication.*

*H7b. The degree of **transparency** in the decision-making process of cobots is correlated with their robustness. A more transparent decision-making process is expected to lead to a higher level of **robustness**, as it allows for better monitoring and understanding of the cobots' actions, enabling them to adapt and handle various situations more effectively.*

*H7c. The **transparency** of the decision-making process in cobots is correlated with their **accuracy**. When cobots' decision-making processes are transparent, their actions and outputs are more aligned with the intended tasks, resulting in higher accuracy.*

*H7d and H7e. The **transparency** of the decision-making process in cobots is significantly related to their **privacy** and **security** measures. When the decision-making process is transparent, it is easier to assess the privacy implications and security vulnerabilities in cobots' operations.*

*H8. The **communication** style of cobots directly influences workers' understanding of **training** content and significantly impacts the effectiveness of the received trainings.*

*H8a and H8b. The communication of cobots with users has a significant correlation with privacy and security. A strong correlation suggests that clear and informative communication fosters a greater sense of trust and confidence in the privacy and security practices implemented in cobots.*

*H9. The **initiator's approach** has a direct effect on workers' **knowledge** regarding cobots. The initiator's efforts to disseminate information, provide training, and raise awareness about cobots are expected to result in a higher level of trust among workers regarding the these cobots.*

*H10. The **initiator's approach has** a direct effect on workers' **hands-on experience** with cobots. By facilitating practical interactions and opportunities to work directly with cobots, the initiator can foster trust and confidence in these cobots among workers.*

*H11. The level of **accuracy** exhibited by cobots directly influences their overall **reliability**. A higher degree of accuracy is expected to result in a more reliable performance of cobots, as their actions and outputs align closely with the intended tasks and objectives.*

*H11a. There is a positive correlation between the **accuracy** and **robustness** of cobots. This means that when cobots are more precise in their actions and outputs, they are likely to exhibit a higher level of adaptability in various operating conditions.*

*H12. Ensuring a high level of **privacy** protection in cobots directly influences their **reliability**. By safeguarding sensitive data and user information, cobots are less susceptible to privacy breaches and potential data manipulations that could compromise their functioning.*

*H12a. The **security** and **privacy** levels of data collected by cobots from job sites are significantly correlated.*

*H13. The level of **security** measures implemented in cobots directly impacts their **reliability**. When cobots are equipped with robust security features, they are better protected against potential cyber threats and unauthorized access.*

The path diagram (Figure 2) summarizes these hypotheses and adheres to conventions regarding the representation and labeling of variables. One-headed arrows signify hypothesized causal relationships, pointing from the cause to the effect. When variables are only correlated without assumed causal connections, a double-headed, curved arrow is used. It is important to account for residual error in predictions, representing the impact of unmeasured predictors in the model. In Structural Equation Models (SEMs), these error terms, referred to as disturbances, are depicted by arrows (sometimes with dotted lines). These disturbances are identified with numbered subscripts ($e_1$, $e_2$, $e_3$, etc.) to indicate their involvement in predicting values for specific variables [27].

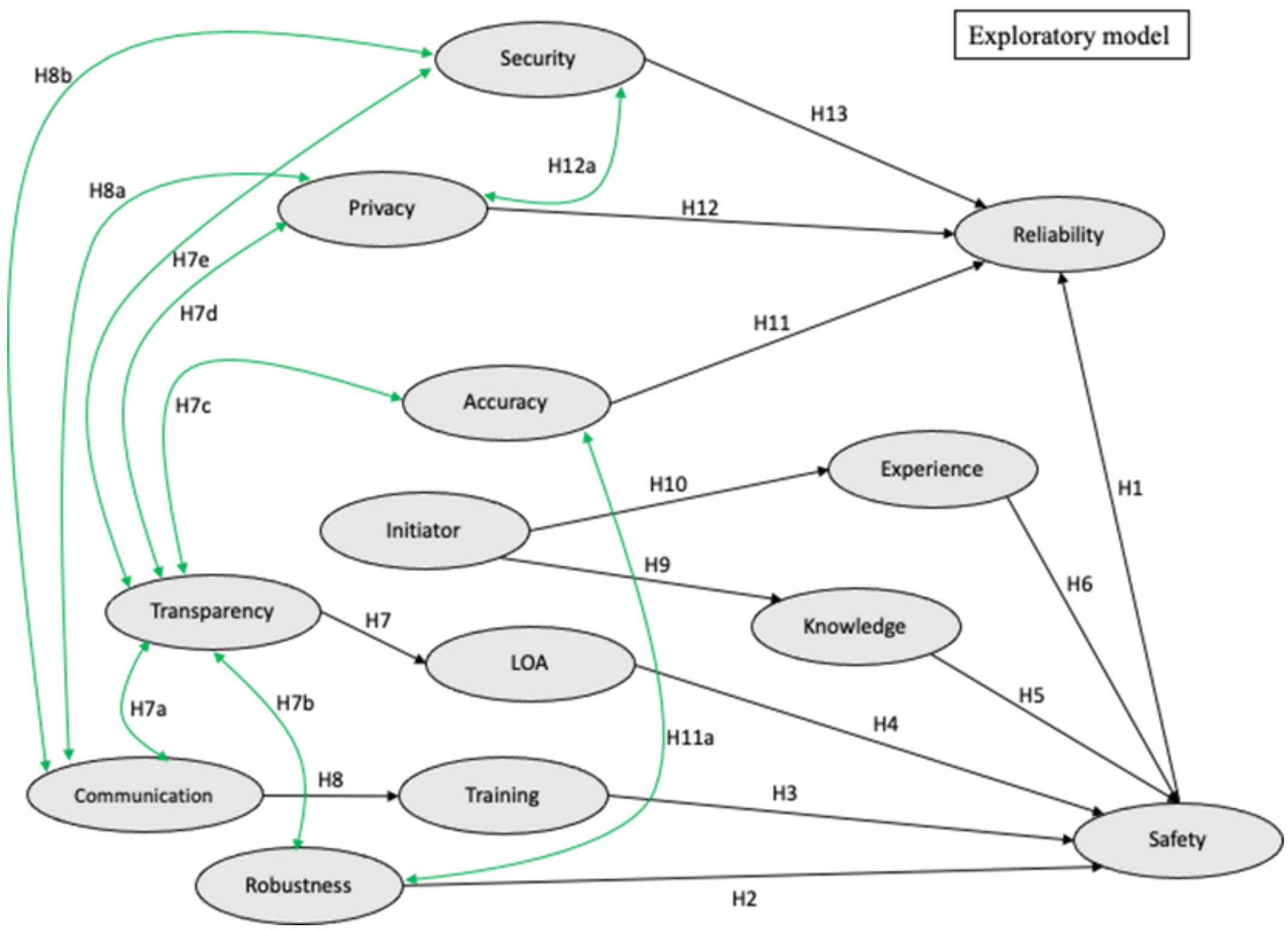

Figure 2. Proposed exploratory model.

## 2. Methodology

### 2.1 Procedure

To accomplish the goals outlined in the preceding section, this research employs a Qualtrics [28] survey to individuals who met at least one of the following criteria: working in the AEC sector, having relevant technology experience in the AEC industry, or having conducted research related to the integration of intelligent cobots/robots in the AEC industry. The survey was distributed using two methods: posting on LinkedIn and recruiting participants through the Prolific survey panel (an online participant platform) [29]. Prolific respondents were prescreened for their employment in the AEC industry.

The necessary sample size for conducting an SEM analysis, the methodology utilized in this study, relies on various elements. For the present study, the online statistics calculator developed by Daniel Soper was employed to estimate the necessary sample size. This calculator takes into account factors such as the expected effect size, desired statistical power level, the number of latent variables, the number of observed variables, and the probability level [30]. A power analysis was carried out to calculate the minimum sample size needed to detect the anticipated effect size of 0.3, with a desired statistical power level of 0.8, a probability level of 0.05, and considering the study's 13 latent variables and 80 observed variables (detailed equations available upon request from the corresponding author). The results indicated that a minimum of 204 participants would be needed to detect the effect, and at least 225 participants would be necessary for model structure testing. Therefore, a total of 300 participants were chosen to perform each of the two SEM analyses, which aligns with the sample size used in prior successful studies on trust towards technology integration in construction [31, 32].

Those who completed the survey via the Prolific survey panel received compensation of $12 per hour, while participants who joined through other channels had a chance to win a $100 gift card. The incentive amounts were chosen carefully: insufficient incentives could lead to low participation and disinterested participants, leading to a biased sample. On the other hand, excessive incentives might attract individuals primarily motivated by the reward rather than a genuine interest in the study's topic [33]. The research was approved by San Diego State University's Institutional Review Board (IRB; protocol HS-2022-0258) for adhering to ethical guidelines.

**2.2 Measurement**

The survey (full survey available upon request from the corresponding author) started with 5 multiple-choice demographic questions (listed in Table 1). Afterwards, participants were given the opportunity to watch a short video demonstrating seven construction cobots in action, aimed at improving their understanding of what cobots are and how they might be used in the construction industry. Following that, there were 28 Likert-scale questions (listed in Table 2) to gauge participants' opinions about factors that might have impact on their level of trust in cobots. Additionally, 3 rank-order questions (listed in Table 1) were included to assess preferences for the situations in which cobots are more likely be embraced. In the final section of the survey, participants were requested to rate their assessment of various attributes such as safety, reliability, inner-workings transparency, security, privacy, robustness, and accuracy on a scale of 1 (Strongly Disagree) to 5 (Strongly Agree) for seven different cobot applications.

The questions with varying response scales have been normalized or standardized for consistency. Regarding to the preference questions, each question has been disaggregated into separate observed variables, with labels assigned based on their positions in the ranking. For instance, for question P1 (refer to Table 1) participants were asked to arrange five options in their preferred order. The options in the 1st place were assigned a score of 5, 2nd place received a score of 4, 3rd place was given a score of 3, 4th place received a score of 2, and 5th place got a score of 1. Similarly, to mitigate the impact of outliers, for questions P2 and P3 with three options (instead of five) the options in 1st place were scored as 5, 2nd place as 3, and 3rd place as 1. This conversion facilitates a more standardized and quantifiable assessment of participants' preferences.

The questions concerning the level of education and years of experience in the construction industry were also scored using a standardized scale (refer to questions D2 and D4 in Table 1). For the level of education, the options "High school or below," "Two-year college," "Four-year college," "Master's degree," and "Doctorate" were converted to scores 1, 2, 3, 4, and 5, respectively.

Table 1. Demographic and Preference questions.

| **Demographic questions** | **Possible Choices/Answers** |
|---|---|
| D1 - In which US State are you currently working? | Alabama (1) ... Wyoming (50) |
| D2 - What is your level of education? | 1. High school or below<br>2. Two-year college<br>3. Four-year college<br>4. Master's degree<br>5. Doctorate |
| D3 - What is your job position? | 1. Executive leader (e.g., CEO, COO, CFO, VP)<br>2. Project executive<br>3. Project manager<br>4. Project engineer<br>5. VDC/BIM manager<br>6. Architect/Designer<br>7. Superintendent<br>8. Foreman<br>9. Laborer<br>10. Other: |
| D4 - How many years of experience do you have in construction? | 1. 1-5<br>2. 6-10<br>3. 11-15<br>4. 16-20<br>5. More than 20 |
| D5 - Which of the listed technologies have you seen used in the projects you were involved in? | 1. Drones<br>2. Robots<br>3. Unmanned tools and machinery (e.g., remotely controlled heavy equipment)<br>4. Other: |
| **Preference questions (Drag and drop choices in a desired order)** | **Possible Choices/Answers** |
| P1 - Rank the following areas in order of your preference for cobots to help improve construction procedures. | 1. Safety<br>2. Productivity<br>3. Quality<br>4. Project cost<br>5. Project duration |
| P2 - Rank the following training types in order of your preference to be more effective for workers in utilizing an AI-powered cobot in your project? | 1. Demonstration videos<br>2. Simulations (e.g., games)<br>3. Real-world hands-on |
| P3 - Which would you prefer? an AI-powered cobot that … | …. understands your work and reacts to it accordingly rather than one that takes orders from you.<br>…. takes orders from you rather than understands your work and reacts to it accordingly.<br>… both understands your work and reacts to it accordingly as well as takes orders from you. |

*Table 2.* Descriptive analysis for Likert questions *(Please read each statement and rate them from 1 (Strongly Disagree) to 5 (Strongly Agree).)*

| # | Statement | Mean | SD |
|---|---|---|---|
| 1 | I have read or heard about the use of robots in the construction industry. | 3.72 | 1.25 |
| 2 | I have directly used a robot in a construction project. | 1.87 | 1.28 |
| 3 | The widespread adoption of AI-powered cobots in construction requires them to be fully trusted first. | 4.16 | 1 |
| 4 | I will be more likely to embrace an AI-powered cobot if its use is recommended by my peers. | 3.75 | 1.04 |
| 5 | I will be more likely to embrace an AI-powered cobot if its use is recommended by my company. | 3.7 | 1.07 |
| 6 | I will be more likely to embrace an AI-powered cobot if its use is recommended by my union. | 3.45 | 1.22 |
| 7 | I will be more likely to embrace an AI-powered cobot if its use is recommended by someone who has directly worked with it. | 4.14 | 0.93 |
| 8 | I will be more likely to embrace an AI-powered cobot if I have had some hands-on experience with it myself. | 4.41 | 0.87 |
| 9 | I will be more likely to embrace an AI-powered cobot if its decision-making process is transparent to me. | 4.2 | 0.94 |
| 10 | I am willing to use an AI-powered cobot if its decision-making process is easy to be understood. | 4.25 | 0.91 |
| 11 | The physical appearance of an AI-powered cobot is an important factor in gaining my trust to use it. | 2.8 | 1.35 |
| 12 | Being able to have a two-way conversation with an AI-powered cobot is an important factor in gaining my trust to use it. | 3.09 | 1.23 |
| 13 | A humanoid AI-powered cobot (a robot that looks like a human), is more trustworthy. | 2.36 | 1.19 |
| 14 | I would prefer to work with a small-sized cobot (left [Reference] ) rather than a big one (right [Reference] ). | 3.44 | 1.21 |
| 15 | I do not trust working with an AI-powered cobot since I think it may get out of control and harm me. | 2.46 | 1.17 |
| 16 | A humanoid AI-powered cobot (a robot that looks like a human), is scary and not likeable. | 2.7 | 1.31 |
| 17 | I will be likely to trust an AI-powered cobot to do some work completely automated and without any human supervision. | 2.95 | 1.13 |
| 18 | I will be likely to trust an AI-powered cobot to do some work only if it is under human supervision. | 3.72 | 1.01 |
| 19 | I will be likely to trust an AI-powered cobot to do some work only if does what I say. | 3.61 | 0.97 |
| 20 | I am willing to adopt an AI-powered cobot if its cost of purchasing and maintaining is not excessively high. | 3.85 | 1.04 |
| 21 | I will be more likely to embrace an AI-powered cobot if it comes with a responsive and responsible customer support service. | 4.04 | 1.02 |
| 22 | I will be more likely to embrace an AI-powered cobot if it comes with a training package. | 4.29 | 0.92 |
| 23 | I will be more likely to embrace an AI-powered cobot if I will be ensured that it leads to upskilled workers who perform higher-level roles. | 3.78 | 1 |
| 24 | I will be more likely to embrace an AI-powered cobot that has been used in well-stablished/well-known construction companies. | 4.06 | 0.98 |
| 25 | I will be more likely to embrace an AI-powered cobot with global/national standards for implementing them. | 3.92 | 1.06 |
| 26 | I will be more likely to embrace an AI-powered cobot with advanced privacy and security systems. | 4.08 | 0.98 |
| 27 | Whether or not the AI-powered cobot can perceive the environment and the extent of this perception will impact my level of trust in it. | 3.84 | 0.99 |
| 28 | If I notice a problem or error in my hands-on experience using an AI-powered cobot, I will be less likely to adopt the cobot on the next try. | 3.58 | 1.04 |

Similarly, for years of experience in the construction industry, the options "1-5," "6-10," "11-15," "16-20," and "More than 20" were converted to scores 1, 2, 3, 4, and 5, respectively. Regarding question D5, the scoring process was as follows: If participants selected all three options, their response was converted to a score of 5. If they chose robots along with either drones or unmanned machinery, their response was converted to a score of 4. Choosing only robots resulted in a score of 3. Opting for drones, unmanned machinery, or both led to a score of 2. Lastly, responses indicating "none" or "others" were assigned a score of 1. This scoring approach was implemented because the research emphasizes robotics, and therefore, responses involving robots were given higher scores to reflect the focus on this aspect.

Reliability pertains to the consistency of results obtained from a measurement instrument, while validity relates to how accurately a measure or instrument assesses the intended construct. Table 3 presents the results of various indices, including Construct Reliability (CR), Average Variance Extracted (AVE), Maximum Shared Squared Variance (MSV), and Average Shared Squared Variance (ASV), along with their acceptable ranges. It is demonstrated in the table that all the measures fall within the acceptable range, indicating satisfactory reliability and validity.

*Table 3.* Reliability and validity assessment.

| Model | Exploratory | | | | Confirmatory | | | |
|---|---|---|---|---|---|---|---|---|
| Latent variable | CR | AVE | MSV | ASV | CR | AVE | MSV | ASV |
| Knowledge | 0.922 | 0.646 | 0.591 | 0.117 | 0.843 | 0.745 | 0.614 | 0.137 |
| LOA | 0.710 | 0.678 | 0.082 | 0.010 | 0.712 | 0.891 | 0.136 | 0.021 |
| Transparency | 0.885 | 0.786 | 0.743 | 0.157 | 0.928 | 0.732 | 0.705 | 0.145 |
| Communication | 0.741 | 0.535 | 0.255 | 0.052 | 0.758 | 0.659 | 0.284 | 0.089 |
| Initiator | 0.703 | 1.658 | 0.931 | 0.145 | 0.701 | 0.597 | 0.469 | 0.121 |
| Privacy | 0.708 | 0.624 | 0.443 | 0.163 | 0.721 | 0.652 | 0.471 | 0.134 |
| Robustness | 0.910 | 0.709 | 0.637 | 0.104 | 0.848 | 0.508 | 0.352 | 0.096 |
| Experience | 1.678 | 3.397 | 0.931 | 0.169 | 1.017 | 0.932 | 0.904 | 0.166 |
| Security | 0.828 | 0.580 | 0.511 | 0.102 | 0.830 | 0.564 | 0.552 | 0.098 |
| Accuracy | 1.014 | 1.039 | 0.937 | 0.104 | 1.027 | 0.767 | 0.716 | 0.115 |
| Acceptable range | > 0.7 | >0.5 | <AVE | <AVE | > 0.7 | >0.5 | <AVE | <AVE |

The researchers employed Structural Equation Modeling (SEM), a statistical technique that enables researchers to investigate and test hypotheses concerning the connections between observed and unobserved (i.e., latent) variables [34]. Unlike other statistical approaches that focus on a single dependent variable and a set of independent variables, SEM permits the modeling of intricate relationships between latent (unobserved) and observed variables [35]. Latent variables are theoretical constructs that cannot be directly measured but are believed to influence attitudes and behaviors and are defined by multiple observed variables [36]. SEM is a valuable tool, particularly in research areas involving complex and multifaceted constructs, like psychology [37].

To evaluate the validity of the measurement constructs in the model, a confirmatory factor analysis (CFA) was performed. CFA is a statistical technique within structural equation modeling that examines the extent to which a group of observed variables (indicators) effectively measures a set of latent variables (factors) that are presumed to exist based on theory or previous studies [38]. The outcomes of the CFA provide insights into whether the items truly capture the same underlying construct. After conducting the CFA, any observed variables with a factor loading below 0.5 were omitted from the analysis. Specifically, five observed variables from question P1, with factor loadings approximately around 0.2, were excluded from further consideration in the analysis.

## 3. Results

### 3.1 Descriptive analyses

These results (see Table 2) indicate the participants' perceptions and attitudes towards adopting AI-powered cobots in the construction industry. On average, the participants indicated moderate awareness of cobots (item #1), but relatively few had direct experience with robots in construction projects (item #2). The participants' perceptions and attitudes toward adopting AI-powered cobots in the construction industry were evident from their responses. Trust emerged as a crucial factor, showing strong agreement among participants (item #3). Additionally, peer recommendations held significant weight in embracing AI-powered cobots, closely followed by company recommendations (item #4&5). Union recommendations seemed to have a slightly lower impact in comparison (item #6).

On average, participants highly valued direct experience with the cobot, emphasizing hands-on interaction (item #8). They also stressed the importance of transparency in the cobot's decision-making process and appreciated understanding this process, while valuing recommendations from those who had worked directly with the cobot (item #7, 9, and 10). Participants had a neutral perception of the cobot's physical appearance (item #11), showing moderate interest in two-way conversations (item #12), but a mixed response to a humanoid appearance (item #13 and 16). Participants displayed a modest preference for smaller cobots, while variability in responses suggests varying opinions (item #14). They also generally expressed reservations about trusting AI-powered cobots, indicating some safety concerns without extreme mistrust, as responses showed variability (item #15). Regarding trust in cobot autonomy, participants were somewhat hesitant to fully trust autonomous cobots (item #17) and they preferred cobots to be under human supervision (item #18). Participants on average rated the importance of the cost of purchasing and maintaining the cobot as significant for adoption (item #20). They also on average, highly valued responsive and responsible customer support services (item #21).

Similarly, on average, participants showed a strong inclination to embrace cobots that included a training package (item #22). Participants on average rate the desire for cobots leading to upskilled workers moderately high (item #23). Recommendations from well-established construction companies are also considered important, with participants expressing a relatively high rating for this factor (item #24). The majority of participants (77.66%) ranked "Real-world hands-on" training as their top choice, indicating a strong preference for practical, hands-on experience with the cobots in real-world settings. "Demonstration videos" were ranked second by 35.58% of participants, suggesting that visual demonstrations and instructional videos are also valued as effective training methods. Meanwhile, "Simulations (e.g., games)" were the second choice for 50.39% of participants, indicating a preference for interactive and gamified training experiences.

Participants on average considered having global/national standards for implementing cobots as essential, and emphasized the importance of advanced privacy and security systems (item #25 and 26). The majority of them believed that the cobot's ability to perceive the environment impacted their level of trust (item #27). However, most of them indicated that if they encountered problems or errors during hands-on experience, they were less likely # 28).

**3.2 Exploratory model**

The proposed model (Figure 2) was tested using a randomly selected 300 participants from the overall sample which included a total of 600 participants. Figure 3 illustrates the results of the proposed model test. The standardized regression coefficients or path coefficient for each path are represented by the numbers on the arrows, which reflect both the magnitude and direction of the relationships between the variables. There are many indices for determining whether the data fit the model, and it is important to consider multiple fit indices rather than relying on a single index. In the next paragraph, recommended cutoffs are described and the results for the exploratory model test for the current study will be indicated in parentheses.

The minimum discrepancy divided by its degrees of freedom (CMIN/df) should typically be below 3 (2.38). The goodness-of-fit index (GFI) is often considered acceptable when around 0.90 or higher (0.75), while the adjusted goodness-of-fit index (AGFI) is expected to fall within the range of 0.85 or higher (0.78). The root mean square error of approximation (RMSEA) is generally considered acceptable when below 0.08 (0.06). Incremental fit indices, such as the comparative fit index (CFI), incremental fit index (IFI), Tucker-Lewis index (TLI), and the relative fit index (RFI), are considered indicative of a good fit when above 0.90 (CFI 0.87, IFI 0.87, TLI 0.86, RFI 0.73). Parsimonious fit indices, such as the normed fit index (NFI), the parsimonious normed fit index (PNFI) and the parsimonious comparative fit index (PCFI), should ideally be around 0.50 or higher (NFI 0.75, PNFI 0.53, PCFI 0.65) [39].

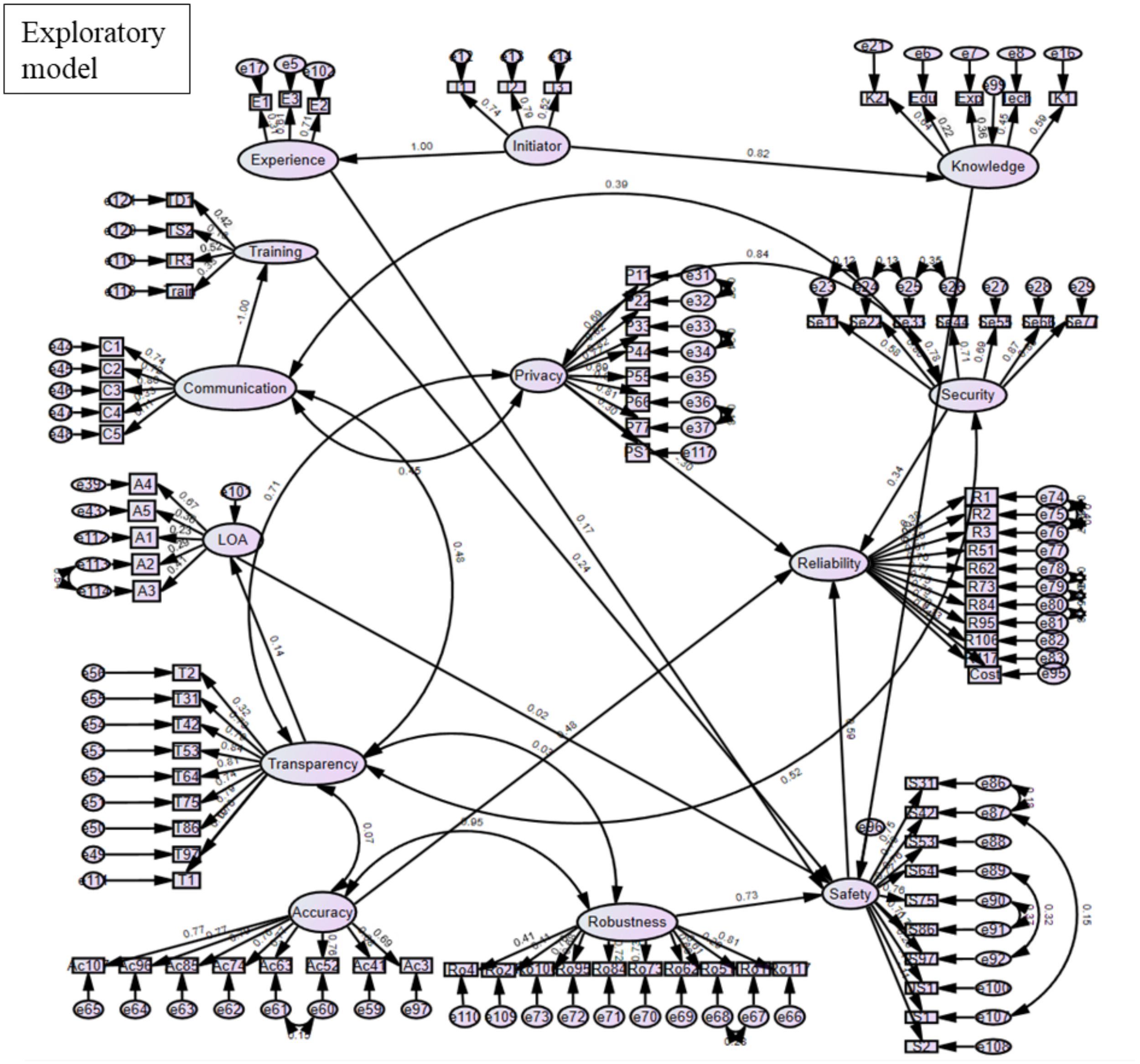

Figure 3. Exploratory model results

Because all incremental fit indices, GFI, and AGFI scores were below the acceptable threshold, it was concluded that the proposed model did not fit well. Consequently, the exploratory model was revised in an attempt to achieve a better fit. The modification indices were employed to incorporate these relationships, but only those that had a logical basis and were supported by prior research and interview data were included. This means that the added correlations were in line with existing literature and interview findings, as suggested by Cho et al. [40].

The results indicated that hypotheses H4, H6, H7, H8, H7b, and H7c were rejected due to their respective p-values being 0.649, 0.212, 0.079, 0.058, 0.611, and 0.145. These findings suggest that there were no significant effects or correlations observed for these hypotheses. Following the utilization of the modification indices and further exploration of the literature, a few hypotheses were incorporated into the model:

*H14. Maintaining a high level of **privacy** in cobot operations directly influences their **safety**.*

*H15. The level of **security** measures implemented in cobots has a direct impact on their **safety**.*

*H16. The **reliability** of cobots has a direct impact on the level of **automation** that can be achieved. More reliable cobots are better equipped to perform tasks consistently and accurately, enabling higher levels of automation with increased confidence in their capabilities.*

*H17. The level of **experience with cobots** directly influences the degree of automation that individuals and organizations are willing to implement. Greater experience and familiarity with cobots lead to increased trust in their abilities, which, in turn, facilitates higher levels of automation.*

*H7f. There is a significant correlation between the **initiator of cobot usage** and the level of **transparency** provided in the cobots' operations. A transparent approach to cobot functionalities, guided by the initiator's initiatives, is expected to contribute to better understanding and acceptance of cobots by users and organizations.*

*H8c. There is a significant correlation between the **communication** style of cobots and the effectiveness of **training** sessions. It suggests that effective and clear communication between cobots and users during training enhances the learning experience and improves the overall effectiveness of the training process.*

*H8d. There is a significant correlation between the **communication** style of cobots and the party responsible for **initiating their usage**.*

*H11b. There is a significant correlation between the* ***accuracy*** *of cobots and the effectiveness of* ***training*** *provided to users. It suggests that cobots with higher accuracy are more likely to facilitate successful and impactful training sessions for users.*

### 3.3 Confirmatory model

The revised model (Figure 4) was tested using the remaining 300 participants from the overall sample.

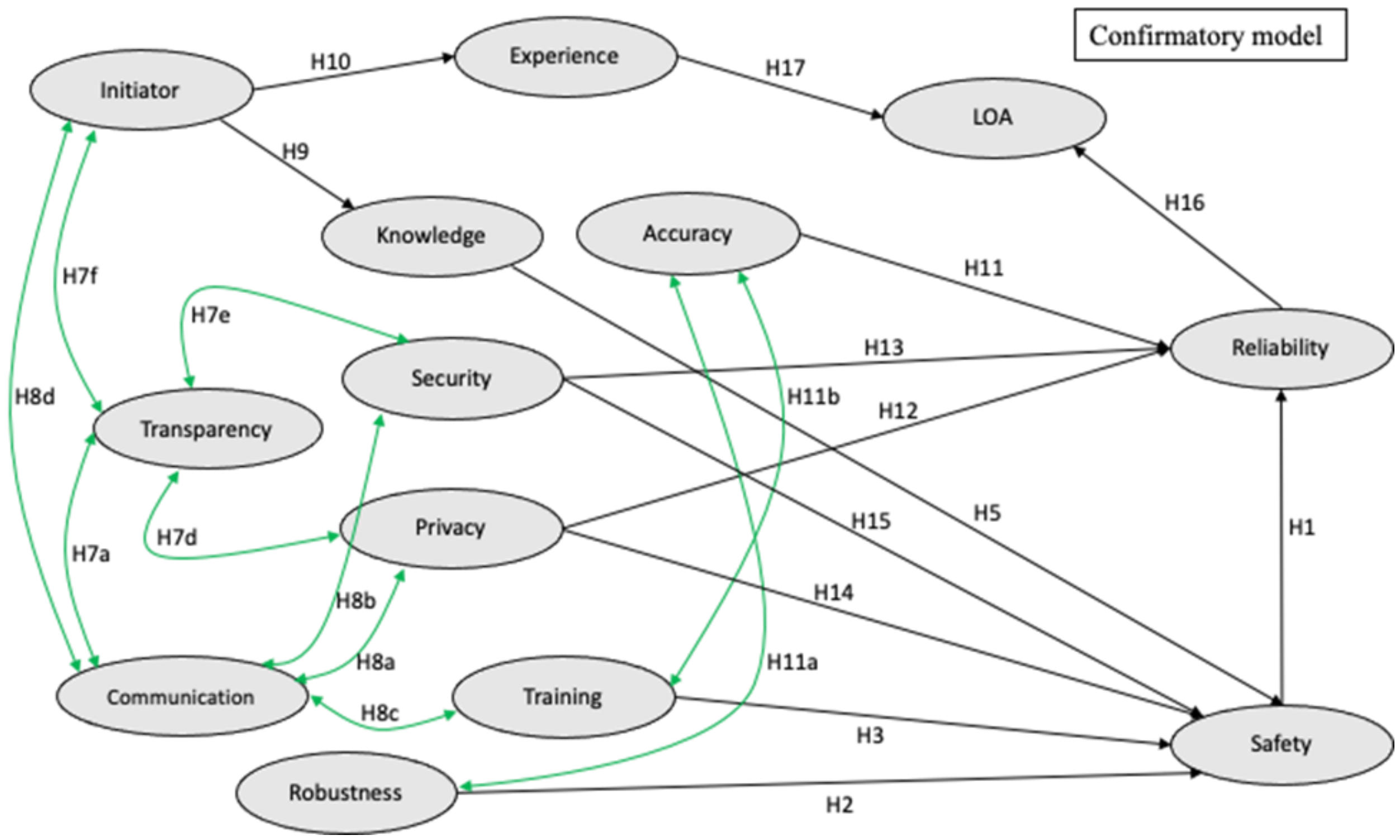

Figure 4. Proposed revised (confirmatory) model.

This validation process aimed to verify the relationships and assumptions proposed in the model using an independent set of data. Figure 5 illustrates the results of the proposed model test. Again, the recommendations for fit indices are described below, with observed values in parentheses. The minimum discrepancy divided by its degrees of freedom (CMIN/df) should typically be below 3 (2.20). The goodness-of-fit index (GFI) is often considered acceptable when around 0.90 or higher (0.84), while the adjusted goodness-of-fit index (AGFI) is expected to fall within the range of 0.85 or higher (0.82). The root mean square error of approximation (RMSEA) is generally considered acceptable when below 0.08 (0.05). Incremental fit indices, such as the comparative fit index (CFI), incremental fit index (IFI), Tucker-Lewis index (TLI), and the relative fit index (RFI), are considered indicative of a good fit when above 0.90 (CFI 0.90, IFI 0.93, TLI 0.91, RFI 0.83). Parsimonious fit indices, such as the normed fit index (NFI), the parsimonious normed fit index (PNFI) and the parsimonious comparative fit index (PCFI), should ideally be around 0.50 or higher (NFI 0.84, PNFI 0.82, PCFI 0.84) [39]. Although the GFI and AGFI are still below the acceptable range, the incremental fit indices are largely above the recommended value, indicating the data fit the model reasonably well.

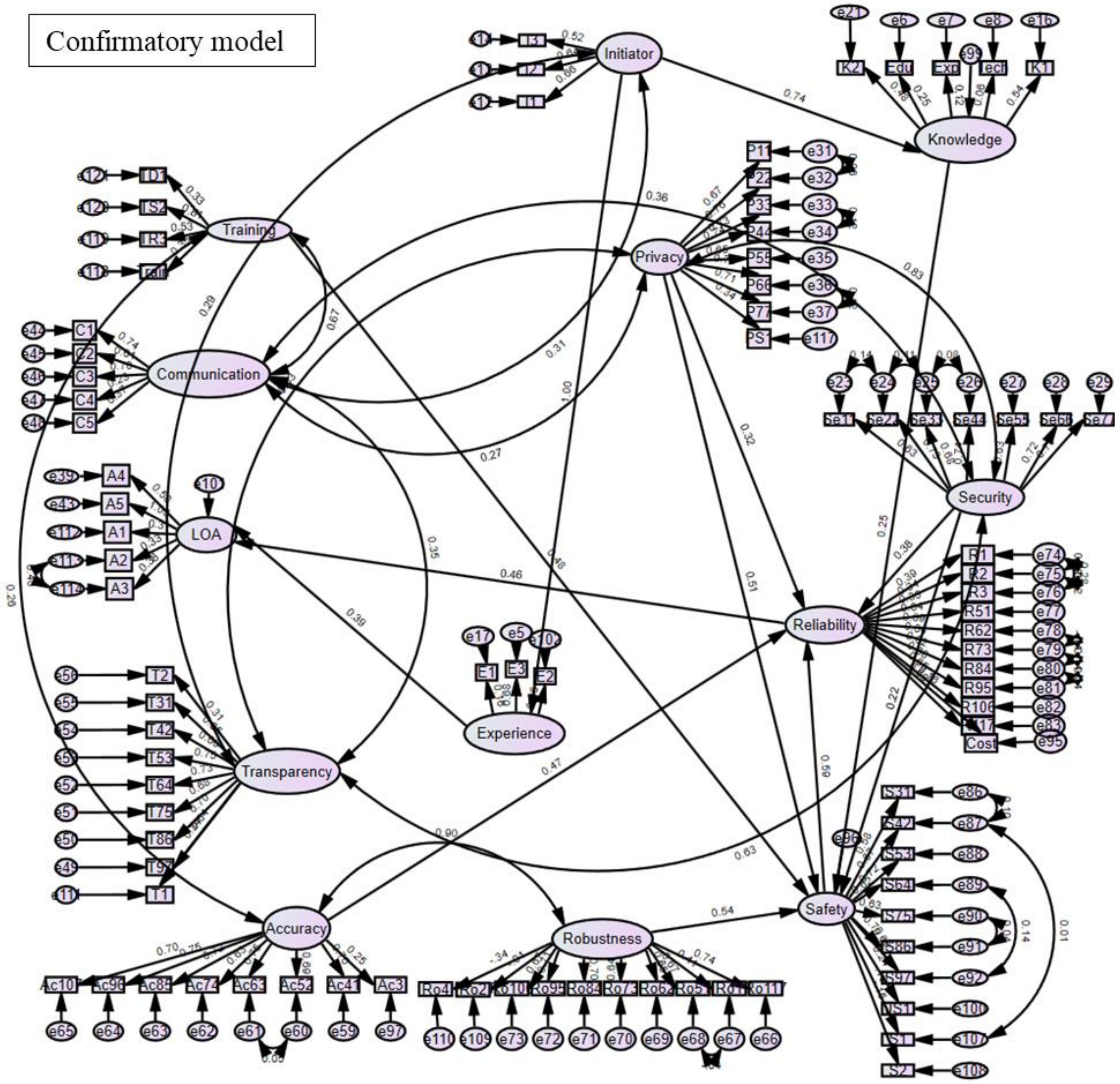

Figure 5. Confirmatory model results.

### 3.4 The perceptions about seven construction cobots in action

A descriptive analysis was conducted on seven demonstrated cobot applications in AEC projects, as mentioned in the previous section. These robots are as follows: The HRP-5P humanoid robot specializes in heavy labor and autonomous operations in hazardous areas, focusing on gypsum board installation [41]. The SAM100 (Semi-Automated Mason) lays bricks at an impressive rate, enhancing worker safety and efficiency [42]. TyBot, an automated steel-tying robot, employs computer vision for rapid installation on construction sites [43]. The robust Husqvarna DXR 305 demolition robot precisely handles demanding tasks in both industrial and construction settings [44]. Boston Dynamics' Spot robot autonomously navigates construction sites, capturing comprehensive 360º images for project documentation [45]. Okibo's autonomous wall plastering technology combines AI, 3D scanners, and sensors to efficiently apply coatings on walls [46]. ERO Concrete Recycling Robot contributes to sustainable construction by safely demolishing concrete buildings for recycling, reducing waste in the process [47]. The purpose was to determine whether the approach to establishing trust should vary depending on the specific cobot application. The analysis also aimed to identify which dimensions are perceived as more crucial from the participants' perspective in overall cobot trust. It is worth mentioning that the standard errors for all attributes in each cobots were implied a similar level of consensus among the participants in their expectations. The relatively low standard error (~0.03) suggests a high level of agreement among the participants regarding the cobot's attributes. Table 4 shows the descriptive statistics on the survey participants perceptions about these cobots.

*Table 4.* Descriptive analysis for the seven demonstrated cobots.

| HRP-5P | | | | | | | |
|---|---|---|---|---|---|---|---|
| | *Safety* | *Reliability* | *Transparency* | *Security* | *Privacy* | *Robustness* | *Accuracy* |
| Mean | 4.0544 | 4.0177 | 3.5063 | 3.8342 | 3.7418 | 3.9127 | 4.1063 |
| Standard Error | 0.0371 | 0.0341 | 0.0381 | 0.0372 | 0.0373 | 0.0348 | 0.0333 |
| Median | 4 | 4 | 4 | 4 | 4 | 4 | 4 |
| Mode | 5 | 5 | 4 | 4 | 4 | 4 | 5 |
| SD | 1.0432 | 0.9591 | 1.0701 | 1.0448 | 1.0497 | 0.9769 | 0.9372 |
| SAM100 | | | | | | | |
| Mean | 4.0051 | 4.0177 | 3.6051 | 3.7367 | 3.6759 | 3.8152 | 4.0987 |
| Standard Error | 0.0368 | 0.0341 | 0.0372 | 0.0379 | 0.0383 | 0.0370 | 0.0324 |
| Median | 4 | 4 | 4 | 4 | 4 | 4 | 4 |
| Mode | 5 | 5 | 4 | 4 | 3 | 4 | 5 |
| SD | 1.0330 | 0.9591 | 1.0445 | 1.0652 | 1.0757 | 1.0398 | 0.9120 |
| TYBOT | | | | | | | |
| Mean | 3.9570 | 3.9544 | 3.5709 | 3.7266 | 3.6620 | 3.8203 | 4.1152 |
| Standard Error | 0.0368 | 0.0353 | 0.0384 | 0.0385 | 0.0397 | 0.0361 | 0.0321 |
| Median | 4 | 4 | 4 | 4 | 4 | 4 | 4 |
| Mode | 5 | 4 | 4 | 4 | 3 | 4 | 5 |
| SD | 1.0334 | 0.9920 | 1.0796 | 1.0816 | 1.1172 | 1.0148 | 0.9010 |
| DXR-305 | | | | | | | |
| Mean | 3.9481 | 3.9823 | 3.6443 | 3.7253 | 3.6810 | 3.9342 | 4.0709 |
| Standard Error | 0.0378 | 0.0343 | 0.0378 | 0.0375 | 0.0386 | 0.0357 | 0.0323 |
| Median | 4 | 4 | 4 | 4 | 4 | 4 | 4 |
| Mode | 5 | 4 | 4 | 4 | 4 | 5 | 5 |
| SD | 1.0638 | 0.9630 | 1.0615 | 1.0545 | 1.0843 | 1.0035 | 0.9077 |
| Spot | | | | | | | |
| Mean | 3.9519 | 3.9405 | 3.4810 | 3.7570 | 3.7392 | 3.8823 | 4.0443 |
| Standard Error | 0.0369 | 0.0352 | 0.0395 | 0.0385 | 0.0385 | 0.0365 | 0.0335 |
| Median | 4 | 4 | 3 | 4 | 4 | 4 | 4 |
| Mode | 5 | 5 | 3 | 5 | 4 | 4 | 5 |
| SD | 1.0368 | 0.9906 | 1.1106 | 1.0818 | 1.0835 | 1.0257 | 0.9402 |
| Okibo | | | | | | | |
| Mean | 3.9253 | 3.9747 | 3.5722 | 3.6797 | 3.7013 | 3.8886 | 4.0848 |
| Standard Error | 0.0373 | 0.0353 | 0.0383 | 0.0388 | 0.0380 | 0.0352 | 0.0326 |
| Median | 4 | 4 | 4 | 4 | 4 | 4 | 4 |
| Mode | 5 | 5 | 4 | 4 | 4 | 4 | 5 |
| SD | 1.0492 | 0.9914 | 1.0771 | 1.0903 | 1.0677 | 0.9880 | 0.9155 |
| ERO | | | | | | | |
| Mean | 3.9190 | 3.9367 | 3.5873 | 3.7253 | 3.6519 | 3.9165 | 4.0835 |
| Standard Error | 0.0390 | 0.0353 | 0.0384 | 0.0372 | 0.0386 | 0.0350 | 0.0319 |
| Median | 4 | 4 | 4 | 4 | 4 | 4 | 4 |
| Mode | 5 | 5 | 3 | 4 | 3 | 4 | 5 |
| SD | 1.0955 | 0.9935 | 1.0795 | 1.0461 | 1.0841 | 0.9843 | 0.8953 |

Across the seven cobots, different attributes seem to hold varying levels of importance. HRP-5P and SAM100 prioritize safety and accuracy, with reliability also valued highly. Transparency received lower ratings for both. TyBot values accuracy the most, followed by reliability, while safety received a slightly lower rating. For DXR-305, robustness and transparency are crucial, while safety and security received slightly lower ratings. Spot values safety and accuracy but rated transparency and privacy lower. Robustness and accuracy are highly valued by Okibo, while safety and security received relatively lower ratings. ERO places significant importance on robustness and accuracy, but safety, reliability, and privacy received lower ratings. The cobot type and application likely influence the perceived importance of different attributes, with safety and accuracy consistently considered vital, while other attributes may vary based on the cobot's specific functionality and intended use.

## 5. Conclusion

The study's findings revealed several significant factors that influence trusting cobots by construction workers. Specifically, this study highlighted the role of perceived safety and reliability of AI-powered cobots towards establishing trust. Training and robustness were identified as key direct contributors to safety perceptions, indicating that providing workers with adequate training and ensuring that cobots can effectively perceive their environment will foster a sense of safety and confidence in their operation. The study also emphasized the importance of proper communication between workers and cobots, which positively affected accuracy, robustness, security, privacy, and the level of automation in the conducted analysis. Additionally, the research highlighted the role of initiators in influencing trust, with their knowledge and experience impacting users' understanding and perceptions of cobots. Recommendations from peers, companies, and individuals who have worked with cobots were also proved influential in shaping their trust. Participants showed some hesitancy in fully trusting fully autonomous cobots, preferring them to operate under human supervision. The results underscored the importance of perceived reliability in gaining trust in cobots, with factors like accuracy and security playing crucial roles. The cost of purchasing and maintaining cobots, along with the availability of responsive customer support and training packages, were also significant factors impacting their willingness to adopt the technology.

Transparency was also found to be significant in building trust, but its importance varied across different cobot applications. Participants placed a high importance on trust in cobots, emphasizing the need for transparent decision-making processes and the value of hands-on experience with the technology. Privacy and security were ranked as essential factors, emphasizing the need for robust measures to safeguard data and ensure a reliable and trustworthy cobot system. The study revealed that direct experience and hands-on involvement with cobots positively influenced trust, leading to increased knowledge and confidence in their capabilities. The provision of comprehensive training, particularly through simulations and real-world experiences, was seen as essential for encouraging cobot acceptance and adoption. In conclusion, this research provides valuable insights into the multifaceted factors that contribute to trust in AI-powered cobots within the construction industry. By understanding and addressing these factors, stakeholders can create a conducive environment for successful cobot integration, fostering user confidence and maximizing the potential benefits of this advanced technology.

**5.1 Limitations**

Although this study represents an important endeavor to explore the factors influencing the adoption of cobots in construction from a trust-building perspective, it does have certain limitations. First, while attempts have been made to involve a diverse range of construction practitioners, such as project laborers, foremen, engineers, managers, and leadership, it is important to examine the distinct viewpoints of each group individually. The introduction of cobots into construction projects can have diverse implications for these stakeholders, and conducting separate analyses of their perspectives can help identify specific challenges or concerns. Therefore, it is vital to investigate the unique outlooks of these various parties in order to gain a comprehensive understanding of the impact of cobots in the construction industry. Second, there was no empirical study conducted by researchers to assess trust in cobots in either controlled experimental environments or real-world field settings where a cobot is deployed. Therefore, conducting real-world investigations to examine the trust dimensions identified in this research, as well as those validated through interviews and surveys, would provide additional confirmation regarding the reliability and validity of the research findings. Third, the level of trust can be influenced by the type and size of construction projects. The type of project determines the complexity and range of tasks that cobots are expected to perform. In smaller construction projects, cobots may handle simpler responsibilities like material transportation or basic assembly. Conversely, larger projects may require cobots to engage in more intricate activities such as welding, cutting, and drilling. The complexity of these tasks can impact workers' trust in cobots, as they may be hesitant to rely on them for more demanding duties. In smaller projects, the presence of a cobot may disrupt the workflow and draw more attention, leading to increased wariness and reduced trust from workers. However, in larger projects with multiple workers and machinery, cobots may integrate more smoothly into the workflow and be more widely accepted. Additionally, the project size can influence the extent of training and supervision provided to workers collaborating with cobots. Smaller projects may prioritize less training and supervision, which could increase the risk of accidents and errors, thus diminishing workers' confidence in the safety of cobots. Conversely, larger projects may emphasize comprehensive training and supervision to minimize risks, thereby enhancing overall trust in cobots.

**5.2 Future Work**

Besides addressing the limitations outlined earlier, the future directions of this work encompasses specific research efforts that currently pursued by the research team. Measurement of specific trust factors can be accomplished using physiological data collected from workers who are involved in construction robotics activities. Furthermore, future research can explore the factors that affect over- and under-trust to establish procedural requirements that result in proper trust calibration. Pilot or case study projects that incorporate real or simulated cobots and actively working with AI-enabled cobots can better help in understanding the system (i.e., cobot), user, and environmental characteristic that influence trust. This approach allows for a more comprehensive assessment of the worker-robot collaboration requirements, considering factors such as potential malfunctions or the need for maintenance that may arise over time. Furthermore, long-term utilization of cobots enables a more precise and realistic evaluation of their capabilities. Hence, the authors suggest longitudinal experiments that allow for long-term worker-robot interaction tests.

## 6. Data Availability Statement

Some or all data, models, or code that support the findings of this study are available from the corresponding author upon reasonable request.

## 7. Acknowledgments

The presented work has been supported by the U.S. National Science Foundation (NSF) CAREER Award through the grant # CMMI 2047138. The authors gratefully acknowledge the support from the NSF. Any opinions, findings, conclusions, and recommendations expressed in this paper are those of the authors and do not necessarily represent those of the NSF.